\documentclass[aps,nofootinbib,prd,reprint,superscriptaddress]{revtex4-1}

\pdfoutput=1
\usepackage{graphicx}

\usepackage{multirow}

\usepackage{amsmath,amssymb,mathrsfs}

\usepackage[bookmarks=false]{hyperref} 
\hypersetup{pdfstartview=FitH,pdfhighlight=/O,colorlinks=false}

\newcommand{\mc}[1]{\mathcal{#1}}

\newcommand{\vp}{V_{\text{pent}}}

\begin{document}

\title{Pentahedral volume, chaos, and quantum gravity}

\author{Hal M. Haggard}
\affiliation{Centre de Physique Th\'eorique de Luminy, Case 907, F-13288 Marseille, EU}
\email{haggard@cpt.univ-mrs.fr}

\begin{abstract}
We show that chaotic classical dynamics associated to the volume of discrete grains of space leads to quantal spectra that are gapped between zero and nonzero volume. This strengthens the connection between spectral discreteness in the quantum geometry of gravity and tame ultraviolet behavior. We complete a detailed analysis of the geometry of a pentahedron,  providing new insights into the volume operator and evidence of classical chaos in the dynamics it generates. These results reveal an unexplored realm of application for chaos in quantum gravity.
\end{abstract}


\maketitle

A remarkable outgrowth of quantum gravity has been the discovery that convex polyhedra can be endowed with a dynamical phase space structure \cite{Bianchi:2011a}. In \cite{Bianchi:2011} this structure was utilized to perform a Bohr-Sommerfeld quantization of the volume of a tetrahedron, yielding a novel route to spatial discreteness and new insights into the spectral properties of discrete grains of space. 

Many approaches to quantum gravity rely on discretization of space or spacetime. This allows one to control, and limit, the number of degrees of freedom of the gravitational field being studied \cite{Rovelli:2010}. In the simplest approaches, such as Regge calculus, attention is often restricted to simplices. However, it is not clear that such a restriction is appropriate for the general study of the gravitational field \cite{Freidel:2010}. The present work takes up the study of grains of space more complex than simplices.

The Bohr-Sommerfeld quantization of \cite{Bianchi:2011} relied on the integrability of the underlying classical volume dynamics, that is, the dynamics generated by taking as Hamiltonian the volume, $H=V_{\text{tet}}$. In general, integrability is a special property of a dynamical system exhibiting a high degree of symmetry. Instead, Hamiltonians with two or more degrees of freedom are generically chaotic \cite{Gutzwiller:1990}. Polyhedra with more than four faces are associated to systems with two or more degrees of freedom and so it is natural to ask: ``Are their volume dynamics chaotic?"

Here we show that the answer to this question has important physical consequences for quantum gravity. Prominent among these is that chaotic volume dynamics implies that there is generically a gap in the volume spectrum separating the zero volume eigenvalue from its nearest neighbors. In loop gravity, it is convenient to work with a polyhedral discretization of space because it allows concrete study of a few degrees of freedom of the gravitational field; however, what is key is the spectral discreteness of the geometrical operators of the theory. This is because the partition functions and transition amplitudes that define the theory are expressed as sums over these area and volume eigenvalues.  The generic presence of gaps (above zero) in the spectra of these operators ensures that these sums will not diverge as smaller and smaller quanta are considered; such a theory should be well behaved in the ultraviolet regime. 

The area spectrum has long been known to be gapped.  However, in the limit of large area eigenvalues, doubts have been raised as to whether there is a volume gap \cite{Brunnemann:2008a}. The detailed study of pentahedral geometry here provides an explanation for and alternative to these results. 

We focus on $H=V$  as a tool to gain insight into the spectrum of volume eigenvalues. The spatial volume also appears as a term in the Hamiltonian constraint of general relativity, the multiplier of the cosmological constant; whether chaotic volume dynamics also plays an important role there is worthy of future investigation.

This work provides two lines of argument: The first line is a detailed study of the classical volume associated to a single pentahedral grain of space. We provide preliminary evidence that the pentahedral volume dynamics is chaotic and all of the analytic tools necessary for a thorough numerical investigation. In particular, we find a new formula for the volume of a pentahedron in terms of its face areas and normals, show that the volume dynamics is adjacency changing, and for the first time analytically solve the Minkowski reconstruction of a non-trivial polyhedron, namely the pentahedron. In the second line general results from random matrix theory are used to argue that a chaotic volume dynamics implies the generic presence of a volume gap.

We initiate the investigation of whether there is chaos in the volume dynamics of gravity by considering a single pentahedral grain of space. As in \cite{Bianchi:2011}, examination of the classical volume dynamics of pentahedra relies on turning the space of convex polyhedra living in Euclidean three-space into a phase space. This is accomplished with the aid of two results: (1) Minkowski's theorem \cite{Minkowski:1897} states that the shape of a polyhedron is completely characterized by the face areas $A_{\ell}$ and face normals $\vec{n}_{\ell}$. More precisely, a convex polyhedron is uniquely determined, up to rotations, by its area vectors $\vec{A}_{\ell} \equiv A_{\ell} \vec{n}_{\ell}$, which satisfy $\sum_{\ell} \vec{A}_{\ell} = 0$, and we call the \emph{space of shapes of polyhedra} with $N$ faces of given areas $A_{\ell}$,
\begin{equation*}
\textstyle \mc{P}_N\equiv \big\{\vec{A}_{\ell},\,\ell=1\,.\,.\,N\, |\;\sum_{\ell} \vec{A}_{\ell}\,=0\,,\,\|\vec{A}_{\ell}\|=A_{\ell}\big\}/SO(3).
\end{equation*}
(2) The space $\mc{P}_N$ naturally carries the structure of a phase space \cite{Kapovich:1996}, with Poisson brackets,
\begin{equation}
\label{eq:PB}
\textstyle \big\{f,g\big\}=\sum_\ell\, \vec{A}_{\ell} \cdot \big(\nabla_{\vec{A}_{\ell}} f \times \nabla_{\vec{A}_{\ell}} g \,\big),
\end{equation}
where $f(\vec{A}_{\ell})$ and $g(\vec{A}_{\ell})$ are functions on $\mc{P}_N$. This is the usual Lie-Poisson bracket if the $\vec{A}_{\ell}$ are interpreted physically as angular momenta, i.e. as generators of rotations.


\begin{figure}[t]
\includegraphics[height=80pt]{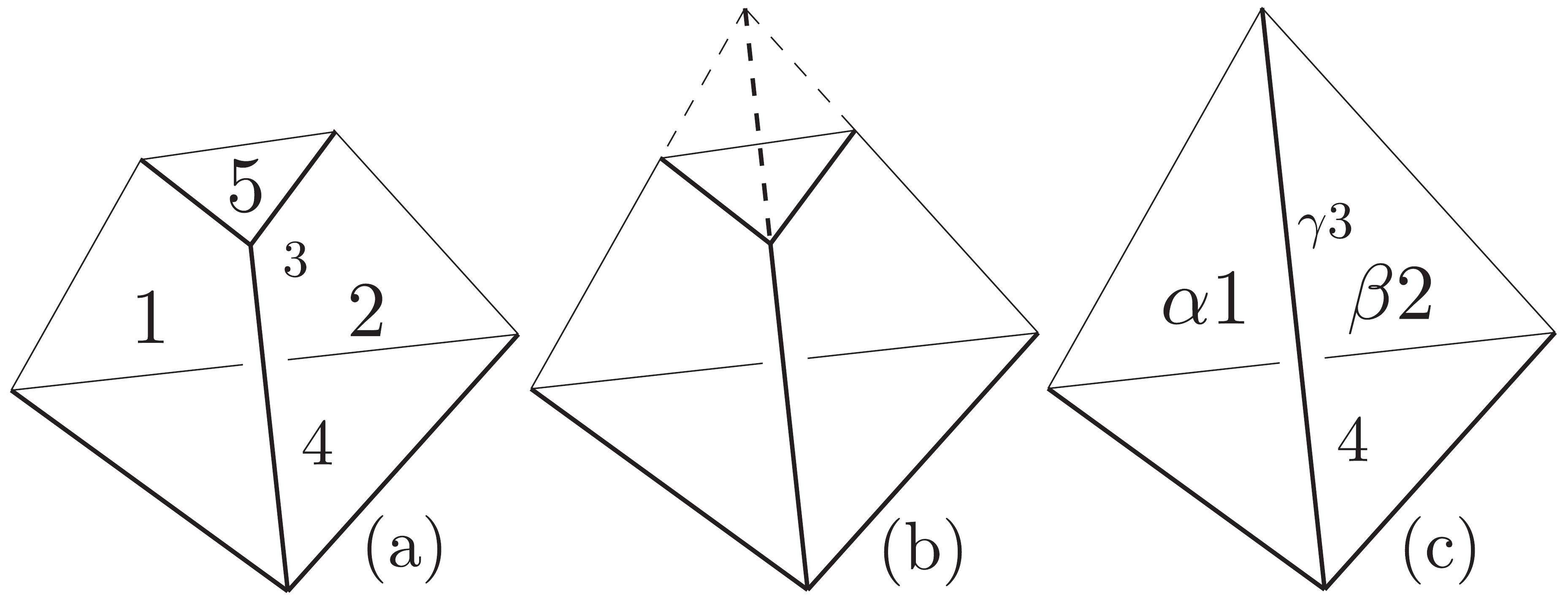}
\caption{Closing a pentahedron into a tetrahedron. (a) A fiducial face labeling of the pentahedron. (b) The closure of the pentahedron upon continuing its sides. (c) The resulting tetrahedron with face labels and associated scalings. }
\label{fig:close}
\end{figure}

To study the pentahedral volume dynamics on $\mc{P}_5$, with $H=V_{\text{pent}}$, it is first necessary to find this volume as a function of the area vectors. Three auxilliary variables, $ \alpha $, $ \beta $, and $ \gamma $,  defined presently, aid in the construction. Assume that a pentahedron with the labels and face adjacencies depicted in Fig. \ref{fig:close}(a) is given. This pentahedron can be closed into a tetrahedron by appropriately continuing the faces 1, 2, and 3; define the positive numbers that scale the old face areas into the new ones to be $ \alpha $, $ \beta $ and $ \gamma $ respectively. Figures \ref{fig:close}(b) and \ref{fig:close}(c) depict the continuation process. Note also that $ \alpha, \beta, \gamma>1$. 

These scalings can be determined using the closure condition of the resulting tetrahedron, $\alpha \vec{A}_{1}+ \beta \vec{A}_2+\gamma \vec{A}_3+\vec{A}_4=0$. By dotting in cross products of any two of $\vec{A}_1$, $\vec{A}_2$, and $\vec{A}_3$ and letting $W_{ijk} = \vec{A}_{i}\cdot(\vec{A}_{j} \times \vec{A}_{k})$ we find,
\begin{equation}
\label{eq:abcDefs}
\alpha = - \frac{W_{234}}{W_{123}}, \quad \beta =  \frac{W_{134}}{W_{123}}, \quad \gamma = - \frac{W_{124}}{W_{123}}.
\end{equation}
Furthermore,  Fig. \ref{fig:close} suggests a formula for the volume of the pentahedron: this volume is the difference of the large tetrahedron's volume (Fig. \ref{fig:close}(c)) and the small dashed tetrahedron's volume (top of Fig. \ref{fig:close}(b)). Thus we have,
\begin{equation}
\label{eq:PentVol}
V_{\text{pent}} = \frac{\sqrt{2}}{3} \left( \sqrt{ \alpha \beta \gamma }-\sqrt{ \bar{\alpha} \bar{ \beta } \bar{ \gamma }} \right) \sqrt{W_{123}},
\end{equation}
where $\bar{\alpha} \equiv \alpha -1$ and similarly for $ \bar{ \beta }$ and $\bar{ \gamma }$. Before investigating the Hamiltonian flow of $V_{\text{pent}}$, it is necessary to examine the role of the fiducial choice made in Fig. \ref{fig:close}(a). 

The Minkowski theorem mentioned above is only an existence and uniqueness theorem. That is, if one is given five vectors satisfying $\sum_{\ell=1}^{5} \vec{A}_{\ell} = 0$, then Minkowski guarantees that a unique pentahedron corresponding to those vectors exists but tells one nothing about how to construct it. As demonstrated, e.g. numerically in \cite{Bianchi:2011a}, the reconstruction problem, i.e. building the polyhedron, is difficult. Lasserre \cite{Lasserre:1983} was the first to appreciate that reconstruction hinges on determining the adjacency of the faces of the end polyhedron from the given vectors. Remarkably, introducing $ \alpha$, $ \beta $, and $ \gamma $ furnishes an analytic solution to the adjacency problem, and subsequently the Minkowski reconstruction for the pentahedron. 

For the dominant class of pentahedra, like that pictured in Fig. \ref{fig:close}(a), there are ten distinct adjacencies and two ``orientations" per adjacency. A convenient method for referring to a pentahedron with a particular adjacency is to state the labeling of the two triangular faces; this determines the adjacencies of all of the faces. By orientation we mean a specification of which triangle is consumed by the continuation process described above. This triangle is referred to as the ``upper" one. (It is tempting to think of it as the ``smaller" triangle but this is not generally true. Rather, the upper triangle can have the larger area, which can be seen by imagining cutting off the tip of Fig. 1 (b) at greater and greater angles with respect to the base giving upper triangles with larger and larger areas.) If we adopt the convention that the first label of a pair denotes the upper triangle we completely set the adjacency and orientation. Thus, Figure \ref{fig:close}(a) depicts a 54-pentahedron. 

While introducing $ \alpha $, $ \beta $, and $ \gamma $ our choice of a 54-pentahedron was fiducial. It is straightforward to list the analogous parameters for each type of pentahedron. For example, a 53-pentahedron with closure $ \alpha^{\prime} \vec{A}_1 + \beta^{\prime} \vec{A}_2 + \vec{A}_3+ \gamma^{\prime} \vec{A}_4=0$ has $ \alpha^{\prime} = W_{234}/W_{124}$, $ \beta^{\prime}  = - W_{134}/W_{124}$ and $ \gamma^{\prime} = -W_{123}/W_{124}$. Remarkably, these can be algebraically expressed in terms of the 54-parameters:
\begin{equation}
\label{eq:CompParams}
\alpha^{\prime} ={ \alpha }/{ \gamma }, \quad \beta^{\prime} = { \beta }/{ \gamma }, \quad \gamma^{\prime} = {1}/{ \gamma } . 
\end{equation}
Now, note that if $ \gamma >1$ then necessarily $ \gamma^{\prime} < 1$ and so the constructability of 54- and 53-pentahedra are mutually exclusive. Thus we see that requiring the closure scalings be greater than one is \textit{a strong condition on constructibility}.  By examining each case similarly, one finds that this condition implies that constructability of 54-pentahedra is only consistent with the constructability of  pentahedra of types $\{12, 21, 23, 32, 13, 31\}$ and exclusive with all other types (see Appendix A).

\begin{figure}[t]
\includegraphics[height=70pt]{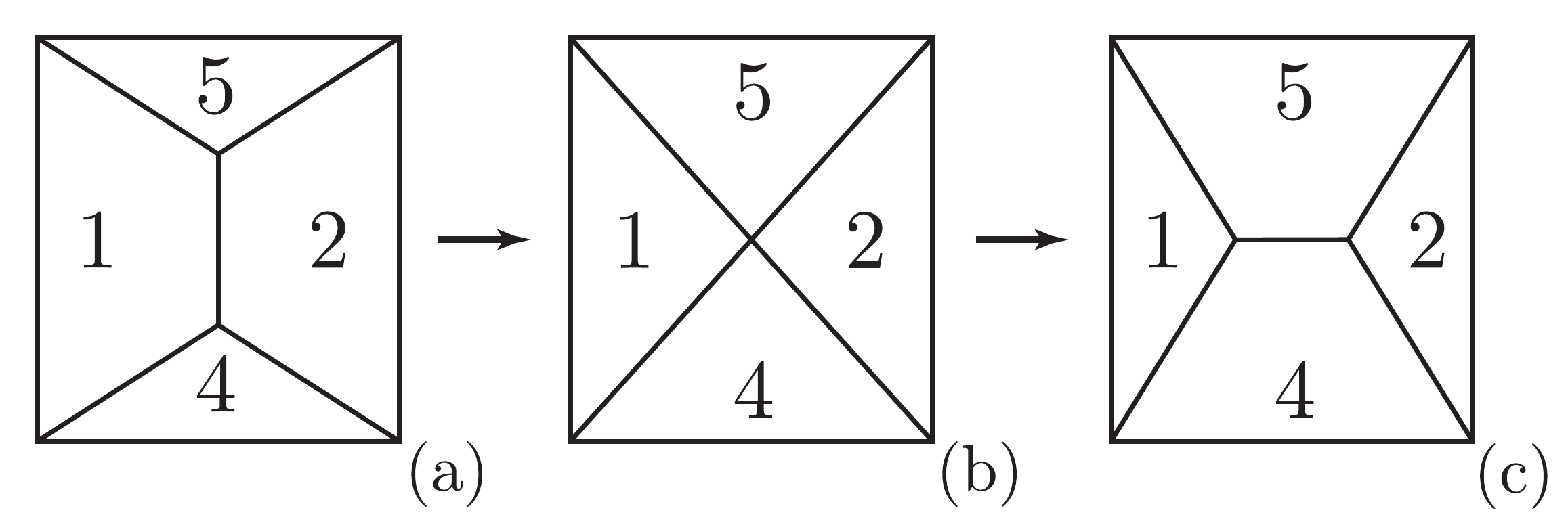}
\caption{Illustrative Pachner move for a pentahedron (the back face, 3, suppressed). (a) The Schlegel diagram of a 54-pentahedron. (b) Schlegel diagram of a sub-dominant quadrilateral pyramid. (c) Resulting Schlegel diagram of a 12-pentahedron upon completion of the  move. }
\label{fig:schlegel}
\end{figure}

These results are more natural when interpreted in light of Schlegel diagrams. A Schlegel diagram is a planar graph that represents a convex polyhedron $P$ by projecting the whole polyhedron into one of its faces. A Pachner move generates a new Schlegel diagram from a given one by contracting an edge until its two vertices meet and then re-expanding a new edge from this juncture in a complementary direction, see Fig. \ref{fig:schlegel}.  Note that the central diagram of Figure \ref{fig:schlegel}(b) is also the Schlegel diagram of a pentahedron, in this case a pyramid built on a quadrilateral base. These are referred to as sub-dominant pentahedra because they are of co-dimension one in the space of all pentahedra; this is due to the fact that four planes meet at the apex of such a pentahedron, a non-generic intersection in three dimensions. We will have more to say about these pyramids briefly. 

The pentahedral types compatible with a 54-pentahedron (listed below \eqref{eq:CompParams}) are precisely those that are reachable by a Pachner move not degenerating a face. The $ \alpha $, $ \beta $, and $ \gamma $ parameters can be used to extend the classification further: ordering the set $\{ \alpha , \beta , \gamma \}$ by magnitude the compatible types can be narrowed to just one, e.g. if $ \alpha > \beta > \gamma $ then only 12 type pentahedra are compatible with type 54. Finally, the last two cases can be distinguished; if $ \gamma \ge \alpha \beta /( \alpha + \beta -1)$ then the type 54 is constructible, else the type 12 is and for equality we have a pyramid, which is a limiting case consistent with both 54- and 12-scalings (see Appendix B). This provides a complete solution to the adjacency problem and leads to analytic formulae for the full Minkowski reconstruction (see Appendix D).

\begin{figure}[t]
\includegraphics[height=170pt]{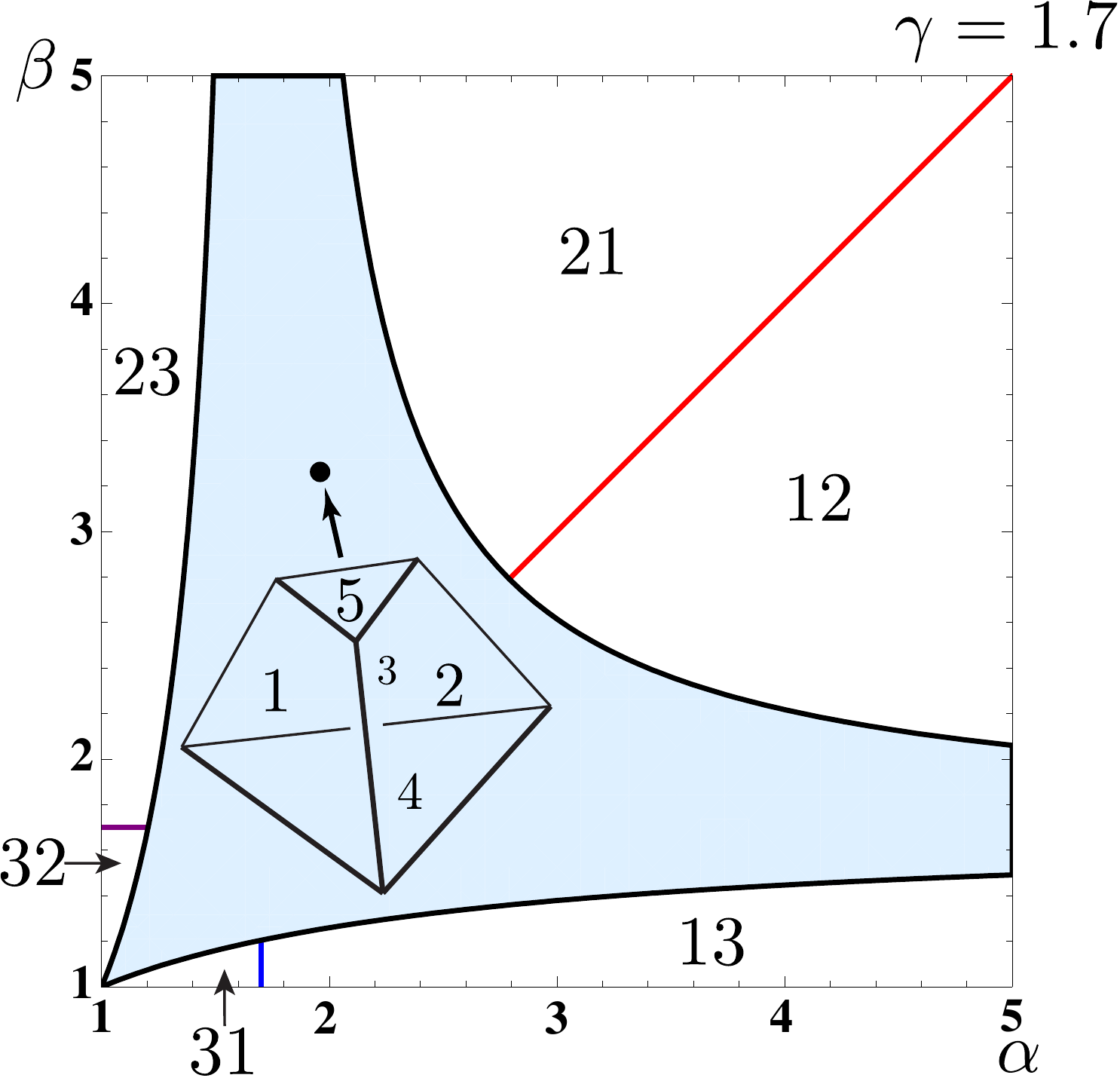}
\caption{A partial phase diagram for the adjacency classes of a pentahedron with fixed $ \gamma =1.7$. The central, shaded region is the parameter space in which the 54-pentahedron is constructible. Neighboring regions are labelled by the pentahedral classes that are constructible within them. }
\label{fig:phase}
\end{figure}

 These findings are summarized in a pentahedral phase diagram, Fig. \ref{fig:phase}. A similar diagram applies with any pentahedral type as the central region, and so, by patching these diagrams together one obtains a phase portrait spanning all pentahedral adjacencies and orientations. 

The importance of all of this is that the formula \eqref{eq:PentVol} only applies to 54-pentahedra. Define $V_{\text{pent}}$ as the function whose level contours consist of all continuously connected equivolume pentahedra in $\mc{P}_5$. We have shown that this function is piecewise defined with a different formula in each adjacency region; each formula obtainable in the same manner as \eqref{eq:PentVol}.  This provides a complete description of $V_{\text{pent}}$ and allows us to begin to address the question of chaos.

The shape space $\mc{P}_5$ has dimension $\dim \mc{P}_5 = 4$, i.e. two degrees of freedom, and again generic Hamiltonians with two degrees of freedom are chaotic \cite{Gutzwiller:1990}. Further, note that the boundaries between the adjacency regions above correspond to quadrilateral pyramids, for example the one of Fig. \ref{fig:schlegel}(b) in the case of a 54/12 boundary. Because the closure scaling works for both pentahedra at a boundary pyramid the volume formulae must agree on the boundary and $V_{\text{pent}}$ is clearly continuous. In fact, see Appendix C, it can be shown that $\vp$ is not smooth but $C^2$, that is, its first and second derivatives are continuous. This surprising result strengthens the expectation that the volume dynamics is chaotic: Dynamical billiards with boundaries that have limited smoothness, an analogous property, frequently exhibit chaos \cite{Chernov:2006}.

The analytic formulae presented in this work make it possible to numerically establish chaos but this will be a lengthy procedure. Here we present early numerical evidence that strengthens the general arguments  given above.  We have numerically implemented the volume dynamics for $H=V_{\text{pent}}$ on the oscillator phase space developed in \cite{Aquilanti:2007}, where the effective tool of symplectic integrators can be applied \cite{Hairer:2002}.  We use an implicit Runge-Kutta symplectic integrator with Gauss coefficients. 
The harmonic oscillator description can be connected, via symplectic reduction, to the shape space $\mc{P}_5$, studied in detail in \cite{Haggard:2010}.  

We can report several interesting pieces of evidence for chaos:  The majority of initial conditions investigated lead to trajectories that cross adjacency boundaries and quickly destabilize the integrator. Adjacency crossing is an entirely new feature, not present (nor possible) in the tetrahedral case. The instability of these trajectories is in accordance with the limited smoothness of $V_{\text{pent}}$ at these boundaries.  By contrast, only a small set of initial conditions give rise to trajectories that do not cross adjacency boundaries. These trajectories can be integrated for long times and Poincar\'e sections created; they correspond to stable tori. Thus the dynamics is certainly of mixed type but instability (chaos) appears to dominate the phase space.  These results are being strengthened by another group \cite{ColemanSmith:2012}. In the special case of equi-area pentahedra, they find evidence of chaos, characterized by positive intermediate Lyapunov exponents, throughout a large fraction of the phase space. The analytic results presented here will allow a complete resolution (for all areas) of the question soon. Therefore it is important to address what chaotic volume dynamics implies about the volume spectrum, which we turn to now. 

\begin{figure}[t]
\includegraphics[height=145pt]{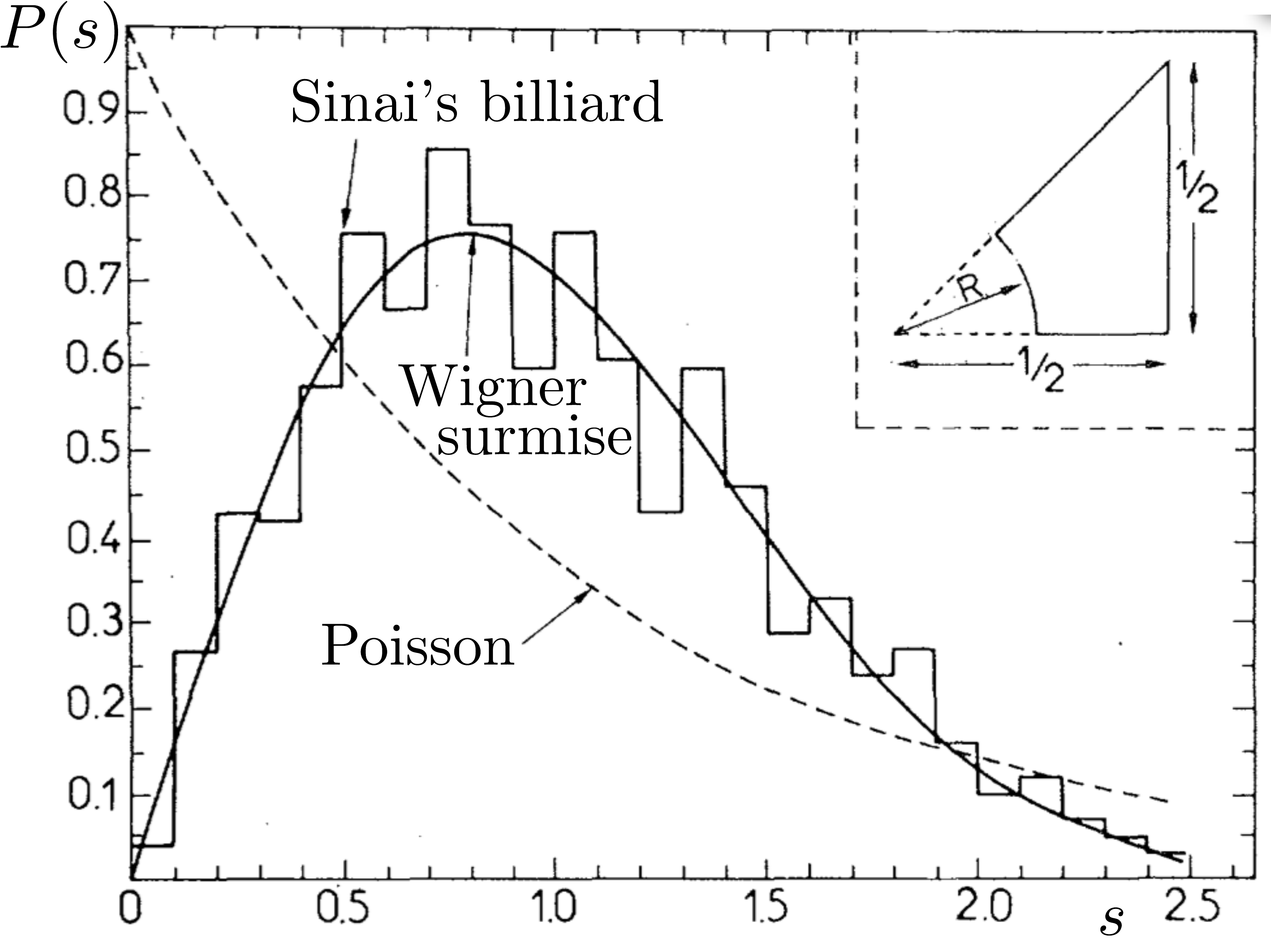}
\caption{The unfolded spectrum for $\sim1000$ consecutive eigenvalues of Sinai's billiard (a prototypical chaotic billiard). The solid (dashed) curve is the prediction for a classically chaotic (integrable) system: Wigner's surmise or Poisson's distribution respectively. Adapted from \cite{Bohigas:1984}.}
\label{fig:LevelStats}
\end{figure}

One route to carry the  integrable-chaotic distinction into quantum theory has been provided by an elegant synthesis of semiclassical and matrix theoretic arguments \cite{Gutzwiller:1990,Haake:2000}. Generic quantum systems with chaotic classical limits have different spectral properties than their classically integrable counterparts. Among these differences is a different expected behavior for the spacing $s$ between neighboring levels, see Fig. \ref{fig:LevelStats}. Classically chaotic systems exhibit level repulsion; at $s=0$, which means degenerate eigenvalues, we see that generically the probability for Wigner's surmise vanishes.  In sharp contrast, classically integrable systems exhibit level bunching, the probability is finite (even maximal) at $s=0$.

To arrive at these results two main steps are taken: First, in order to compare spectra across different systems, the spectra are ``unfolded." Unfolding is a normalization procedure whereby the average level spacing in a given energy interval is brought to one. Once the spectra have been unfolded they can be characterized by their statistical fluctuations, e.g. by the probability of a given level spacing $P(s)$. The second step then is to find $P(s)$ for different types of systems. Random matrix theory has established, for the chaotic case, Wigner's surmise 
\begin{equation}
\label{eq:Wig}
P(s) =\frac{\pi}{2} s \exp(-\pi s^2/4),
\end{equation}
as an excellent approximation for Hermitian Hamiltonian systems having time reversal symmetry.  By contrast semiclassical results give  $P(s) =e^{-s}$ for the integrable case. The generic validity of the connection between random matrices and chaos, originally an empirical observation, has been demonstrated semiclassically \cite{Muller:2009}. 


 It is the level repulsion of the Wigner surmise that guarantees the presence of a volume gap when the classical limit of the volume dynamics is chaotic: if there is a zero volume eigenvalue, repulsion forbids the accumulation of further eigenvalues on top of it. Also, repulsion continues to hold for mixed phase spaces, such as the pentahedral one, with a sufficiently large proportion of chaotic orbits \cite{Haake:2000}.  
 
 The presence of a volume gap in the (integrable) case of a tetrahedron has already been established in \cite{Brunnemann:2008a} and \cite{Bianchi:2011}. Furthermore, a chaotic pentahedral volume dynamics would strongly suggest that there will be chaos for polyhedra with more faces, which have an even richer structure in their phase space \cite{Bianchi:2011a}. Consequently, the argument presented here provides a very general mechanism that would ensure a volume gap for all discrete grains of space. 
 
 One might worry that this result could be spoiled by the unfolding process mentioned above. This concern is justified because more and more states crowd into the interval of allowed volume in the limit of large areas. However, we are interested in the density of states at small volumes (analogously low energies), in fact zero volume, and in this limit the density of states must smoothly go to zero. This is because zero volume polyhedra correspond to collinear configurations of area vectors \cite{Bianchi:2011}, which are individual points of the phase space. These points are not as large as a Planck cell and so cannot support many quantum states. Thus, taking into account the average behavior of the spectrum near zero energy, ``refolding" if you like, may cause some squeezing of states but will not destroy a volume gap. The numerical results of \cite{ColemanSmith:2012} illustrate this general argument in the equi-area case. 

Previous numerical investigations of the quantum volume for 5-valent spin networks (corresponding to the pentahedra here) have, in stark contrast, found evidence of an accumulation at zero volume of the volume eigenvalues \cite{Brunnemann:2008a}. There are multiple reasons for this disagreement. Two important reasons are clear in the present framework: the well studied Ashtekar-Lewandowski and Rovelli-Smolin volume operators \cite{Rovelli:1994ge} both assume that the volume of a region can (i) be broken into a sum of parallelepiped contributions that are of the same form and (ii) these contributions are minimally coupled. The pentahedral volume formula, Eq. \eqref{eq:PentVol}, suggests that (i) is not a good approximation and that there is a strong coupling between the area vectors (fluxes) for more generally shaped regions.  Furthermore, the parallepiped volume is proportional to the tetrahedral one studied in \cite{Bianchi:2011} and there found to be integrable. So the minimal coupling assumption of these proposals may keep the classical dynamics near integrability and lead to the accumulations found in \cite{Brunnemann:2008a}.   The polyhedral volume operator studied here is semiclassically consistent and lucidly exposes geometrical structures, like the strong coupling of Eq. \eqref{eq:PentVol}. 

In this note we have completely solved the geometry of a pentahedron specified by its area vectors and defined its volume as a function of these variables. By performing a numerical integration of the corresponding volume dynamics we have given early indicators that it generates a chaotic flow in phase space. These results uncover a new mechanism for the presence of a volume gap in the spectrum of quantum gravity: the level repulsion of quantum systems corresponding to classically chaotic dynamics. The generic presence of a volume gap further strengthens the expected ultraviolet finiteness of quantum gravity theories built on spectral discreteness.

Thank you to E.~Bianchi for insights and generous encouragement. Also to R.~Littlejohn, A.~Essin and D.~Beke for early interest and ideas. This work was supported by the NSF IRFP grant OISE-1159218. 

\appendix
\section{Table of closure scalings}

This appendix contains an exhaustive list of the closure relations for pentahedra. Each closure relation is also algebraically related to the scaling parameters $ \alpha \equiv \alpha_1$, $ \beta \equiv \beta_1$, and $ \gamma \equiv \gamma_1$ of the first case and once again the shorthand $W_{ijk} \equiv \vec{A}_{i} \cdot (\vec{A}_{j} \times \vec{A}_{k})$ is used. The pentahedron corresponding to each case can be read off by first noting which vector doesn't appear in the closure relation, this vector corresponds to the upper triangle and then noting which vector has no multiplier, this vector corresponds to the lower triangle. Thus case 1. gives the parameters for a 54-pentahedron. 

This list can be used to confirm the claims in the main text about the mutually exclusive constructibility of different adjacency classes. For example, if $ \alpha$, $ \beta $, and $ \gamma $ are all greater than $1$, then necessarily $ \gamma_2$ is less then 1 and the 53-pentahedron is not constructible, see Eq. \eqref{eq:53}. In this manner one can check that a pentahedron whose constructibility is mutually consistent with that of a 54-pentahedron is from the set $\{12, 21, 23, 32, 13, 31\}$. Furthermore, if the parameters $ \alpha $, $ \beta $, and $ \gamma $ are ordered the mutually consistent set can be further narrowed. For example, assume that $ \alpha > \beta > \gamma >1$ then only 54- and 12-pentahedra are mutually consistent; as an illustrative check note that under this assumption $ \alpha_{15} <0$ and so the 23-pentahedron is no longer constructible.  The equations necessary to resolve this final ambiguity (e.g. between 54- and 12-pentahedra) are described in the next section.  \\

\begin{widetext}
\noindent 1. $\alpha_1 \vec{A}_1+ \beta_1 \vec{A}_2 + \gamma_1 \vec{A}_3 +\vec{A}_4 = 0$,
\begin{equation}
\alpha \equiv \alpha_1 = -\frac{W_{234}}{W_{123}} \qquad \beta \equiv \beta_1 = \frac{W_{134}}{W_{123}} \qquad \gamma \equiv \gamma_1 = -\frac{W_{124}}{W_{123}}.
\end{equation}
\noindent 2. $\alpha_2 \vec{A}_1+ \beta_2 \vec{A}_2 + \vec{A}_3 + \gamma_2 \vec{A}_4 = 0$,
\begin{equation}
\label{eq:53}
\alpha_2 = \frac{W_{234}}{W_{124}}=\frac{ \alpha }{ \gamma } \qquad  \beta_2 =- \frac{W_{134}}{W_{124}}=\frac{ \beta }{ \gamma } \qquad \gamma_2 = -\frac{W_{123}}{W_{124}}=\frac{1}{ \gamma }.
\end{equation}
\noindent 3. $\alpha_3 \vec{A}_1+  \vec{A}_2 + \beta_3 \vec{A}_3 + \gamma_3 \vec{A}_4 = 0$,
\begin{equation}
\alpha_3 = -\frac{W_{234}}{W_{134}}=\frac{ \alpha }{ \beta } \qquad  \beta_3 =- \frac{W_{124}}{W_{134}}=\frac{ \gamma }{ \beta } \qquad \gamma_3 = \frac{W_{123}}{W_{134}}=\frac{1}{ \beta }.
\end{equation}
\noindent 4. $ \vec{A}_1+\alpha_4  \vec{A}_2 + \beta_4 \vec{A}_3 + \gamma_4 \vec{A}_4 = 0$,
\begin{equation}
\alpha_4 = -\frac{W_{134}}{W_{234}}=\frac{ \beta }{ \alpha } \qquad  \beta_4 = \frac{W_{124}}{W_{234}}=\frac{ \gamma }{ \alpha } \qquad \gamma_4 = -\frac{W_{123}}{W_{234}}=\frac{1}{ \alpha }.
\end{equation}
\noindent 5. $\alpha_5 \vec{A}_1+ \beta_5 \vec{A}_2 + \gamma_5 \vec{A}_3 +\vec{A}_5 = 0$,
\begin{equation}
 \alpha_5 = -\frac{W_{235}}{W_{123}} =1- \alpha \qquad \beta_5 = \frac{W_{135}}{W_{123}} =1- \beta \qquad  \gamma_5 = -\frac{W_{125}}{W_{123}}= 1- \gamma .
\end{equation}
\noindent 6. $\alpha_6 \vec{A}_1+ \beta_6 \vec{A}_2 +\vec{A}_3 + \gamma_6 \vec{A}_5=0$,
\begin{equation}
\alpha_6 =\frac{W_{235}}{W_{125}} =\frac{1- \alpha}{1- \gamma } \qquad \beta_6 = -\frac{W_{135}}{W_{125}} =\frac{1- \beta}{1- \gamma } \qquad  \gamma_6 = -\frac{W_{123}}{W_{125}}= \frac{1}{1- \gamma}.
\end{equation}
\noindent 7. $\alpha_7 \vec{A}_1+  \vec{A}_2 +\beta_7 \vec{A}_3 + \gamma_7 \vec{A}_5=0$,
\begin{equation}
\alpha_7 =-\frac{W_{235}}{W_{135}} =\frac{1- \alpha}{1- \beta } \qquad \beta_7 = -\frac{W_{125}}{W_{135}} =\frac{1- \gamma }{1- \beta } \qquad  \gamma_7 = \frac{W_{123}}{W_{135}}= \frac{1}{1- \beta } .
\end{equation}
\noindent 8. $ \vec{A}_1+  \alpha_8 \vec{A}_2 +\beta_8 \vec{A}_3 + \gamma_8 \vec{A}_5=0$,
\begin{equation}
\alpha_8 =-\frac{W_{135}}{W_{235}} =\frac{1- \beta }{1- \alpha  } \qquad \beta_8 = \frac{W_{125}}{W_{235}} =\frac{1- \gamma }{1- \alpha } \qquad  \gamma_8 = -\frac{W_{123}}{W_{235}}= \frac{1}{1- \alpha } .
\end{equation}
\noindent 9. $  \alpha_9 \vec{A}_1+\beta_9  \vec{A}_2 +\gamma_9 \vec{A}_4 +  \vec{A}_5=0$,
\begin{equation}
\alpha_9 =-\frac{W_{245}}{W_{124}} =1-\frac{\alpha  }{\gamma  } \qquad \beta_9 = \frac{W_{145}}{W_{124}} =1-\frac{\beta }{\gamma } \qquad  \gamma_9 = -\frac{W_{125}}{W_{124}}=1- \frac{1}{\gamma } .
\end{equation}
\noindent 10. $  \alpha_{10} \vec{A}_1+\beta_{10}  \vec{A}_2 +\vec{A}_4 +\gamma_{10}   \vec{A}_5=0$,
\begin{equation}
\alpha_{10} =\frac{W_{245}}{W_{125}} =\frac{\gamma - \alpha  }{\gamma-1  } \qquad \beta_{10} =- \frac{W_{145}}{W_{125}} =\frac{\gamma - \beta }{\gamma-1 } \qquad  \gamma_{10} = -\frac{W_{124}}{W_{125}}=\frac{\gamma }{\gamma -1} .
\end{equation}
\noindent 11. $  \alpha_{11} \vec{A}_1+ \vec{A}_2 +\beta_{11} \vec{A}_4 +\gamma_{11}   \vec{A}_5=0$,
\begin{equation}
\alpha_{11} =-\frac{W_{245}}{W_{145}} =\frac{\gamma - \alpha  }{\gamma-\beta  } \qquad \beta_{11} =- \frac{W_{125}}{W_{145}} =\frac{\gamma - 1 }{\gamma-\beta } \qquad  \gamma_{11} = \frac{W_{124}}{W_{145}}=\frac{\gamma }{\gamma -\beta} .
\end{equation}
\noindent 12. $  \vec{A}_1+\alpha_{12}  \vec{A}_2 +\beta_{12} \vec{A}_4 +\gamma_{12}   \vec{A}_5=0$,
\begin{equation}
\alpha_{12} =-\frac{W_{145}}{W_{245}} =\frac{\gamma - \beta  }{\gamma-\alpha  } \qquad \beta_{12} = \frac{W_{125}}{W_{245}} =\frac{\gamma - 1 }{\gamma-\alpha } \qquad  \gamma_{12} = -\frac{W_{124}}{W_{245}}=\frac{\gamma }{\gamma -\alpha } .
\end{equation}

\noindent 13. $\alpha_{13}  \vec{A}_1+ \beta_{13} \vec{A}_3 +\gamma_{13} \vec{A}_4 +   \vec{A}_5=0$,
\begin{equation}
\alpha_{13} =-\frac{W_{345}}{W_{134}} =1-\frac{\alpha  }{\beta } \qquad \beta_{13} = \frac{W_{145}}{W_{134}} =1-\frac{\gamma  }{\beta }  \qquad  \gamma_{13} = -\frac{W_{135}}{W_{134}}=1-\frac{1  }{\beta } .
\end{equation}

\noindent 14. $\alpha_{14}  \vec{A}_1+ \beta_{14} \vec{A}_3 + \vec{A}_4 + \gamma_{14}  \vec{A}_5=0$,
\begin{equation}
\alpha_{14} =\frac{W_{345}}{W_{135}} =\frac{\beta -\alpha  }{\beta-1 } \qquad \beta_{14} = -\frac{W_{145}}{W_{135}} =\frac{\beta -\gamma  }{\beta-1 }   \qquad  \gamma_{14} = -\frac{W_{134}}{W_{135}}=\frac{\beta   }{\beta-1 } .
\end{equation}

\noindent 15. $\alpha_{15}  \vec{A}_1+ \vec{A}_3 + \beta_{15} \vec{A}_4 + \gamma_{15}  \vec{A}_5=0$,
\begin{equation}
\alpha_{15} =-\frac{W_{345}}{W_{145}} =\frac{\beta -\alpha  }{\beta-\gamma } \qquad \beta_{15} = -\frac{W_{135}}{W_{145}} =\frac{\beta -1  }{\beta-\gamma }   \qquad  \gamma_{15} = \frac{W_{134}}{W_{145}}=\frac{\beta   }{\beta-\gamma  } .
\end{equation}

\noindent 16. $  \vec{A}_1+ \alpha_{16} \vec{A}_3 + \beta_{16} \vec{A}_4 + \gamma_{16}  \vec{A}_5=0$,
\begin{equation}
\alpha_{16} =-\frac{W_{145}}{W_{345}} =\frac{\beta -\gamma  }{\beta-\alpha } \qquad \beta_{16} = \frac{W_{135}}{W_{345}} =\frac{\beta -1  }{\beta-\alpha }   \qquad  \gamma_{16} = -\frac{W_{134}}{W_{345}}=\frac{\beta   }{\beta-\alpha  } .
\end{equation}

\noindent 17. $ \alpha_{17}  \vec{A}_2+\beta_{17}  \vec{A}_3 +\gamma_{17}  \vec{A}_4 +  \vec{A}_5=0$,
\begin{equation}
\alpha_{17} =-\frac{W_{345}}{W_{234}} =1-\frac{\beta   }{\alpha } \qquad \beta_{17} = \frac{W_{245}}{W_{234}} =1-\frac{\gamma   }{\alpha }   \qquad  \gamma_{17} = -\frac{W_{235}}{W_{234}}=1-\frac{1   }{\alpha }.
\end{equation}

\noindent 18. $ \alpha_{18}  \vec{A}_2+\beta_{18}  \vec{A}_3 +  \vec{A}_4 + \gamma_{18} \vec{A}_5=0$,
\begin{equation}
\alpha_{18} =\frac{W_{345}}{W_{235}} =\frac{\alpha -\beta   }{\alpha-1 } \qquad \beta_{18} =- \frac{W_{245}}{W_{235}} =\frac{\alpha -\gamma   }{\alpha-1 } \qquad  \gamma_{18} = -\frac{W_{234}}{W_{235}}=\frac{\alpha    }{\alpha-1 }.
\end{equation}

\noindent 19. $ \alpha_{19}  \vec{A}_2+  \vec{A}_3 +\beta_{19} \vec{A}_4 +\gamma_{19}  \vec{A}_5=0$,
\begin{equation}
\alpha_{19} =-\frac{W_{345}}{W_{245}} =\frac{\alpha -\beta   }{\alpha-\gamma } \qquad \beta_{19} =- \frac{W_{235}}{W_{245}} =\frac{\alpha -1   }{\alpha-\gamma  } \qquad  \gamma_{19} = \frac{W_{234}}{W_{245}}=\frac{\alpha    }{\alpha-\gamma }.
\end{equation}

\noindent 20. $   \vec{A}_2+\alpha_{20}  \vec{A}_3 +\beta_{20} \vec{A}_4 +\gamma_{20}  \vec{A}_5=0$,
\begin{equation}
\alpha_{20} =-\frac{W_{245}}{W_{345}} =\frac{\alpha -\gamma    }{\alpha- \beta  } \qquad \beta_{20} =\frac{W_{235}}{W_{345}} =\frac{\alpha -1   }{\alpha-\beta  } \qquad  \gamma_{20} = -\frac{W_{234}}{W_{345}}=\frac{\alpha    }{\alpha-\beta }.
\end{equation}

\end{widetext}

\section{Pyramidal pentahedra}

As briefly observed in the main text, the boundary between two adjacency regions consists of pyramidal pentahedra. For these pentahedra both of the closing scalings of the two neighboring adjacency regions are valid and thus the volume of the pentahedral pyramids can be calculated in two distinct manners. Setting these two volumes equal one finds a constraint satisfied amongst the three scaling parameters. For example, at the adjacency between a 54-pentahedron and a 12-pentahedron the following constraint is satisfied,
\begin{equation}
\label{eq:pyramid}
\gamma = \frac{ \alpha \beta }{ \alpha + \beta -1}.
\end{equation}
While the argument outlined above is geometrically obvious the algebraic manipulations are unnecessarily complex. To avoid this complexity we provide another even simpler geometric argument here. 

To fix notations consider the transition from a 54-pentahedron to 12-pentahedron. As discussed in the main text this transition occurs through a Pachner move, see Fig. 3 of the main text. Notice that during this move the edge bordering faces 1 and 2 goes from having non-zero length when the 54-pentahedron is constructible, to having zero length when the pentahedron is pyramidal, and then vanishes altogether when the 12-pentahedron is constructible. This is the observation we will use to derive Eq. \eqref{eq:pyramid}. 

Figure 5 shows the face corresponding to $\vec{A}_1$ of the initial 54-pentahedron, along with the larger, triangular face that appears in Fig. 2(b). Let the edge lengths of the triangle be $\ell_1$, $\ell_2$, and $\ell_3$. Introduce two parameters $ \lambda $ and $ \mu$ that scale the edges $\ell_1$ and $\ell_2$ to give the corresponding edge lengths of the unscaled pentahedron face $ \lambda \ell_1$ and $\mu \ell_2$; these paremeters are necessarily less than one, $ \lambda, \mu<1$. 
\begin{figure}[htbp] 
   \centering
   \includegraphics[width=2.3in]{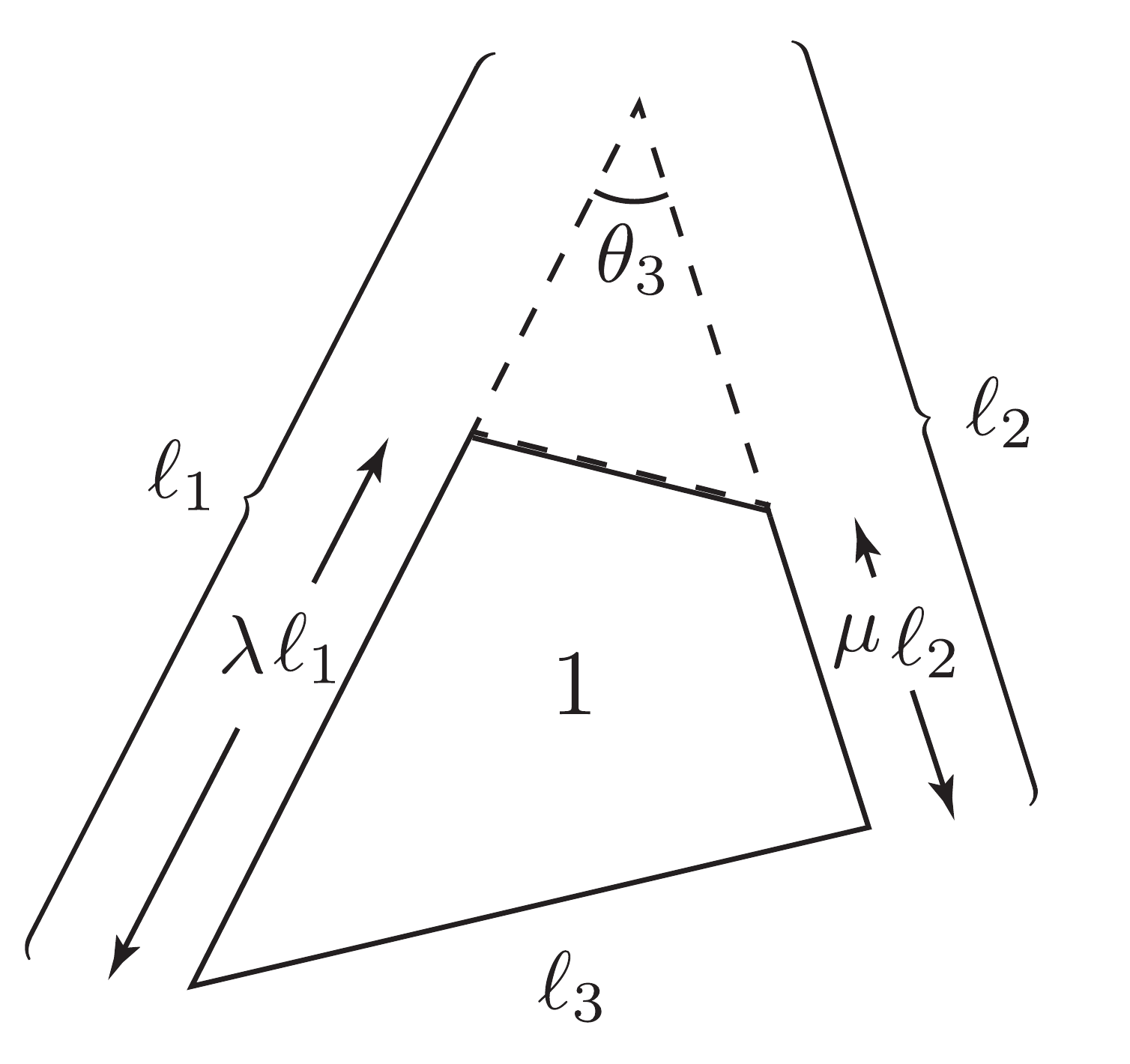} 
   \caption{The first face of a 54-pentahedron and its scaling into a triangle through the closure process described in the main text.  }
   \label{fig:example}
\end{figure}

From the definition of $ \alpha $ the area of the large triangle is $A= \alpha A_1$. On the other hand,  this area can also be expressed in terms of the angle $\theta_3$, $A= 1/2 \ell_1 \ell_2 \sin{\theta_3}$. Similarly, the area of the small dashed triangle is 
\begin{equation}
A_{d} = \frac{1}{2} (1- \lambda )\ell_1 (1- \mu )\ell_2 \sin{\theta_3} = (1- \lambda )(1-\mu) A,
\end{equation}
and this yields a second formula for the area of face 1, the quadrilateral of the diagram,
\begin{equation}
A_1 = A-A_d = A[1-(1- \lambda )(1- \mu )].
\end{equation}
Setting the two expressions for $A_1$ equal yields a relation between the face scalings and the edge scalings:
\begin{equation}
\alpha = \frac{1}{ \lambda + \mu - \lambda \mu }.
\end{equation}
A structurally identical argument applied to the faces 2 and 3 yields the two further relations
\begin{equation}
\beta =  \frac{1}{  \mu+\nu -  \mu \nu }, \qquad \gamma =  \frac{1}{ \nu + \lambda  - \nu \lambda  },
\end{equation}
where $ \nu $ is a final edge scaling used to bring the third edge that meets at the apex of the scaled tetrahedron down to its length in the pentahedron (this is the edge where faces 2 and 3 meet). 

During the Pachner move depicted in Fig. 3, the edge joining faces 2 and 3 degenerates into a point. This can only happen if $\mu=0$ and hence at the pyramidal configuration we have $\alpha =1/ \lambda $ and $ \beta = 1/\nu$. Putting these relations into the expression for $ \gamma $ yields a relation between the area vectors at the 12/54 pyramidal configuration:
\begin{equation}
\label{eq:pyr}
\gamma = \frac{ \alpha \beta }{ \alpha + \beta -1}.
\end{equation}
Furthermore, for the 54-pentahedron to be constructible we must have $\mu>0$ and a straightforward algebraic inversion of the equations above shows that this occurs when $ \gamma $ is greater than the right hand side of Eq.  \eqref{eq:pyr}. These are the results quoted in the main text.

\section{Smoothness of the pentahedral volume}
As explained in the main text $V_{\text{pent}}$ is certainly continuous. However, in order for $V_{\text{pent}}$ to define a Hamiltonian flow it is important to check that at least its first derivatives are continuous as well. It turns out that $V_{\text{pent}}$ is $C^2$, that is, its first and second derivatives are both continuous. 

Because $V_{\text{pent}}$ is defined in a piecewise manner over the different adjacency regions and the formulae in each region are different it is plausible that $V_{\text{pent}}$ is not smooth. To illustrate the demonstration that $V_{\text{pent}}$ is $C^2$ we focus on just two formulae expressing it in the 54 and 12 regions respectively as:
\begin{equation}
 V_{\text{pent}} = 
\begin{cases}
 \frac{\sqrt{2}}{3} \left( \sqrt{ \alpha \beta \gamma }-\sqrt{ (\alpha-1) (\beta -1)(\gamma -1)} \right) \sqrt{W_{123}}, \\
 \frac{\sqrt{2}}{3} \left( \sqrt{\frac{ \alpha  - \gamma }{ \alpha  - \beta  }\frac{ \alpha  -1}{ \alpha  - \beta  }\frac{\alpha }{ \alpha  - \beta  } }-\sqrt{ \frac{\beta - \gamma }{ \alpha  - \beta  } \frac{\beta  -1}{ \alpha - \beta  } \frac{ \beta  }{ \alpha  - \beta   }} \right) \sqrt{W_{345}}, \\
\dots . 
\end{cases}
\end{equation}
It will be easier to work with the function $V_{\text{pent}}/\sqrt{W_{123}}$, which will have the same smoothness as $V_{\text{pent}}$ as long as $W_{123}=0$ is avoided. The advantage of the latter function is that it can be viewed as a function of $ \alpha, \beta,$ and $ \gamma $ alone because $W_{345}/W_{123} = ( \alpha - \beta )$. Again we gloss over consideration of points where these coordinates fail, i.e. where the denominators in their definitions are zero. 

Now, we just check that the first and second derivatives of $V_{\text{pent}}/\sqrt{W_{123}}$ with respect to $ \alpha, \beta,$ and $ \gamma $ and evaluated at $\gamma =  \alpha \beta /( \alpha + \beta -1)$ are the same when calculated in region 54 and in region 12. At third derivatives the two calculations begin to disagree. Finally, this calculation is carried out for each of the boundaries connecting two adjacency regions. 

\section{Minkowski reconstruction}
Having determined the adjacency class of a pentahedron it is straightforward to explicitly reconstruct it. For definiteness, once again, assume that we have established that the adjacency is that of a 54-pentahedron. Choose the origin of coordinates at the vertex where the faces 2, 3, and 4 meet. In these coordinates, only two of the five perpendicular heights to the faces are left to be found as $h_2=h_3=h_4=0$. 

Now, the vector pointing to the intersection of three planes can be expressed in terms of the normals to the planes $\vec{n}_r, \vec{n}_s,$ and $\vec{n}_t$ by,
\begin{equation}
\begin{aligned}
\label{eq:norms}
x_{rst} = \frac{1}{|\vec{n}_r\cdot(\vec{n}_s \times \vec{n}_t)|} [ &h_r(\vec{n}_s\times \vec{n}_t)\\
&+h_t(\vec{n}_r\times \vec{n}_s)+h_s(\vec{n}_{t} \times \vec{n}_r)],
\end{aligned}
\end{equation}
as can be confirmed by dotting in $\vec{n}_r, \vec{n}_s,$ and $\vec{n}_t$.  This formula can be used to find the edge lengths of face 4 and they are all proportional to $h_1$. Let these three edge lengths be $h_1 e_1, h_1 e_2,$ and $h_1 e_3$ where the dimensionless ÒlengthsÓ $(e_1, e_2, e_3)$ are completely determined by the given area vectors and the formula \eqref{eq:norms}. Then using Heron's formula for the area $\Delta$ of a triangle given its edge lengths, $A_4= h_1^2 \Delta(e_1,e_2,e_3)$, and this can be solved for $h_1$,
\begin{equation}
h_1 = \sqrt{\frac{A_4}{\Delta(e_1, e_2, e_3)}}
\end{equation}
Finally, $h_5$ can be extracted from the relation of the volume to the heights, 
\begin{equation}
V_{\text{pent}} = \frac{1}{3} ( A_1 h_1+ A_5 h_5),
\end{equation}
and the value of the volume determined by Eq. (5) of the main text. This completes the Minkowski reconstruction.

\providecommand{\href}[2]{#2}\begingroup\raggedright\endgroup

\end{document}